\documentclass[5p]
{elsarticle}

\usepackage{amsmath}
\usepackage{graphicx}
\usepackage{mathdots}
\usepackage{epstopdf} 
\usepackage[usenames,dvipsnames]{color}
\usepackage{bm}
\usepackage{tabularray}

\usepackage{inputenc}
\usepackage{xcolor}
\usepackage{tikz}

\usepackage[colorlinks,bookmarks=false,citecolor=blue,linkcolor=red,urlcolor=blue]{hyperref}
\usepackage[colorlinks,bookmarks=false,citecolor=blue,linkcolor=red,urlcolor=blue]{hyperref}
\usepackage{xcolor}
\usepackage{mathdots}
\usepackage{float}
\usepackage{needspace}
\usepackage{soul}
\usepackage[usenames,dvipsnames]{color}

\begin{document}
\title{Training iterated protocols for distillation of GHZ states with variational quantum algorithms}

\author[add1,add2]{\'Aron Rozgonyi}

\author[add1,add2]{G\'abor Sz\'echenyi}

\author[add2]{Orsolya K\'alm\'an}

\author[add2]{Tam\'as Kiss}

\address[add1]{ELTE E\"otv\"os Loránd University, H-1117 Budapest, Hungary}

\address[add2]{HUN-REN Wigner Research Centre for Physics, H-1525 Budapest, Hungary}

\begin{abstract}
   We present optimized distillation schemes for preparing Greenberger-Horne-Zeilinger (GHZ) states. Our approach relies on training variational quantum circuits with white noise affected GHZ states as inputs. 
   Optimizing for a single iteration of the scheme, we find that it is possible to achieve an increased fidelity to the GHZ state, although further iterations decrease the fidelity. The same scheme, acting on coherently distorted pure-state inputs, is effective only in certain special cases.  
   We show that radically different results can be achieved, however, when one optimizes for the output after two iterations of the protocol. In this case, the obtained schemes are more effective in distilling GHZ states from inputs affected by white noise. Moreover, they can also correct several types of coherent pure-state errors.

\end{abstract}

\maketitle

\begin{sloppypar}

\section{Introduction}

Multipartite entanglement is fundamentally different from bipartite entanglement \cite{multipartite_review,Szalay} with a number of applications in quantum information protocols, such as conference key agreement \cite{conf_key_review}. In the quest for building large quantum networks, distribution and distillation of genuine tripartite entanglement is a crucial step. The quintessential form of a tripartite entangled state is the GHZ state. Since the pioneering work \cite{Knight_1998} there has been considerable effort devoted to the theory of distilling multipartite entangled states \cite{PhysRevLett.91.107903,PhysRevA.71.012319,Wallnfer,PhysRevResearch.3.033164,Bugalho2023distributing} 
and, especially, GHZ states \cite{PhysRevLett.84.5908,Elkouss,PhysRevResearch.5.023124}. 

The theory of entanglement distillation has produced several important results about asymptotic protocols, where the same type of operations are repeated infinitely many times. Such an approach necessarily requires a rapidly (in general exponentially) growing number of inputs as a function of the number of iterations. Another concept of entanglement distillation, which addresses the impracticality of the requirement of too many inputs, is the one-shot approach, where one aims at optimizing the quality of the output with the restriction of a given number of available inputs. The problem in this case is that brute force optimization efforts become very costly even for a moderate number of inputs  processed simultaneously \cite{Krastanov_1,Krastanov_2}.

Iterations of the simplest circuits, involving entangling operations, measurement(s), and post-selection, can lead to complex dynamical behaviours even for single-qubit systems \cite{Kiss_etal,malachov2019phase}. In the case of two-qubit systems, similar measurement-induced iterated protocols may lead to nontrivial time evolution of the entanglement in the asymptotic limit \cite{kiss2011measurement}.  The latter example clearly indicates that even-odd oscillations can occur in the number of iterations, suggesting that optimization for two iterative steps might reveal properties fundamentally different from one-shot approaches.

Variational Quantum Algorithms (VQAs) are based on the concept of employing classical computers to optimize and train parameterized quantum circuits \cite{vqa-nature, verdon2018universal}.
Exploiting hybrid quantum-classical optimization through parameterized quantum circuits \cite{McClean_2016} is a promising method for obtaining precise solutions for numerical problems when the computational complexity grows exponentially with the size of the system
e.g. calculation of the ground state energy of a given Hamiltonian \cite{VQE-McClean,VQE-Kandala},
learning probability distributions \cite{qGAN-Zoufal},
generating error correction code \cite{johnson2017qvector},
compressing quantum data \cite{Romero_2017},
solving combinatorial problems \cite{farhi2014quantum} 
or learning entanglement distillation \cite{loccnet,chittoor2022learning}.

In this paper, we address the problem of efficiently preparing high fidelity GHZ states by LOCC (Local Operations and Classical Communication) schemes, if we have a source of somewhat distorted GHZ states at hand. 
The protocol is designed by training a parameterized quantum circuit through VQA. 
We are aiming at correcting both coherent distortions of the input and white noise, being the most important type of loss of coherence due to transmission through a depolarizing channel. In order to keep the operations as simple as possible, all parties will use two inputs to operate on simultaneously, and follow the same procedure to distill GHZ states, by applying one or two iterations of the procedure.

\section{Setup}

We possess a state generator that prepares entangled 3-qubit  GHZ states 
\begin{equation}
|\textrm{GHZ}\rangle = \frac{1}{\sqrt{2}} \left(|000\rangle + |111\rangle\right).
\end{equation}
The three qubits are then distributed among three separate parties situated at distant locations. Due to imperfections in both the state generator and the quantum channel, the GHZ state may get distorted. The error can both coherently change the quantum state and reduce its entanglement. 
In what follows, we design an iterative distillation scheme which can correct these errors.

\begin{figure}[h!]
\includegraphics[width=0.45\textwidth]{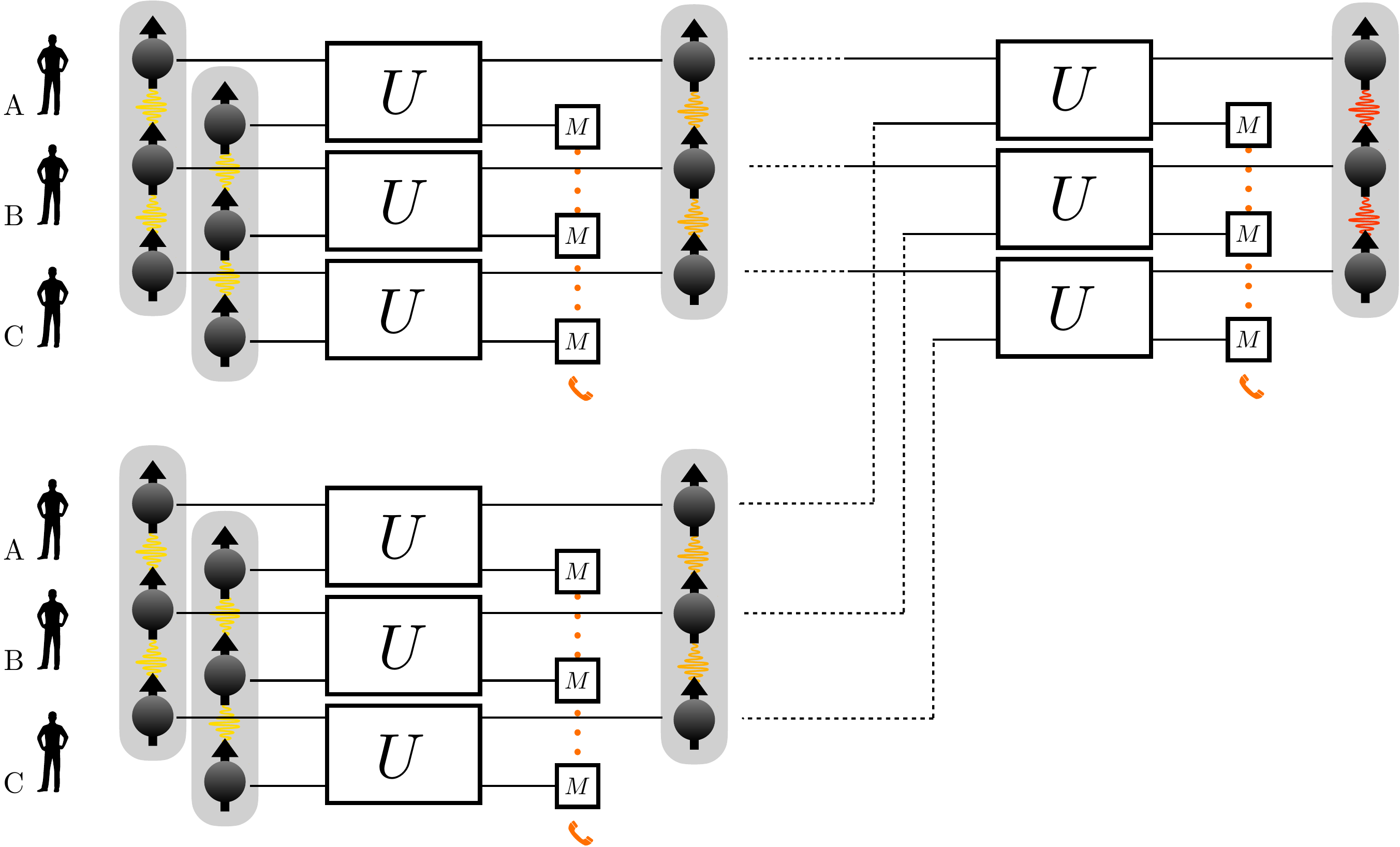}
\caption{\label{fig:setup} A schematic diagram illustrating the nonlinear protocol employed for GHZ distillation, considering a scenario involving two iterations. Three parties share two noisy GHZ states on which they perform local operations (denoted by the unitary operator $U$) and subsequent measurements ($M$). They communicate through classical channels to post-select the states by a consensus process for the next iteration.}
\end{figure}

Figure~\ref{fig:setup} depicts the considered distillation protocol which consists the following steps:
(1) 
The parties initially share two copies of the noisy partially entangled GHZ state. Qubits of one of the copies will be used as flags.
(2) Each party performs identical local unitary operations ($U$) on their qubits.
(3) Each party performs projective measurements in the computational basis ($M$) on the flag qubits.
They only keep the unmeasured copy if all measurement outcomes are zeros, which they can agree on by using classical communication.
If any other measurement outcome is observed, they discard the unmeasured copy and repeat the steps (1)-(3).

Our purpose is to optimize the entangling operation $U$ so that few iterations of the protocol result in an output with high fidelity to the GHZ state.

\begin{figure}[]
\includegraphics[width=0.45\textwidth]{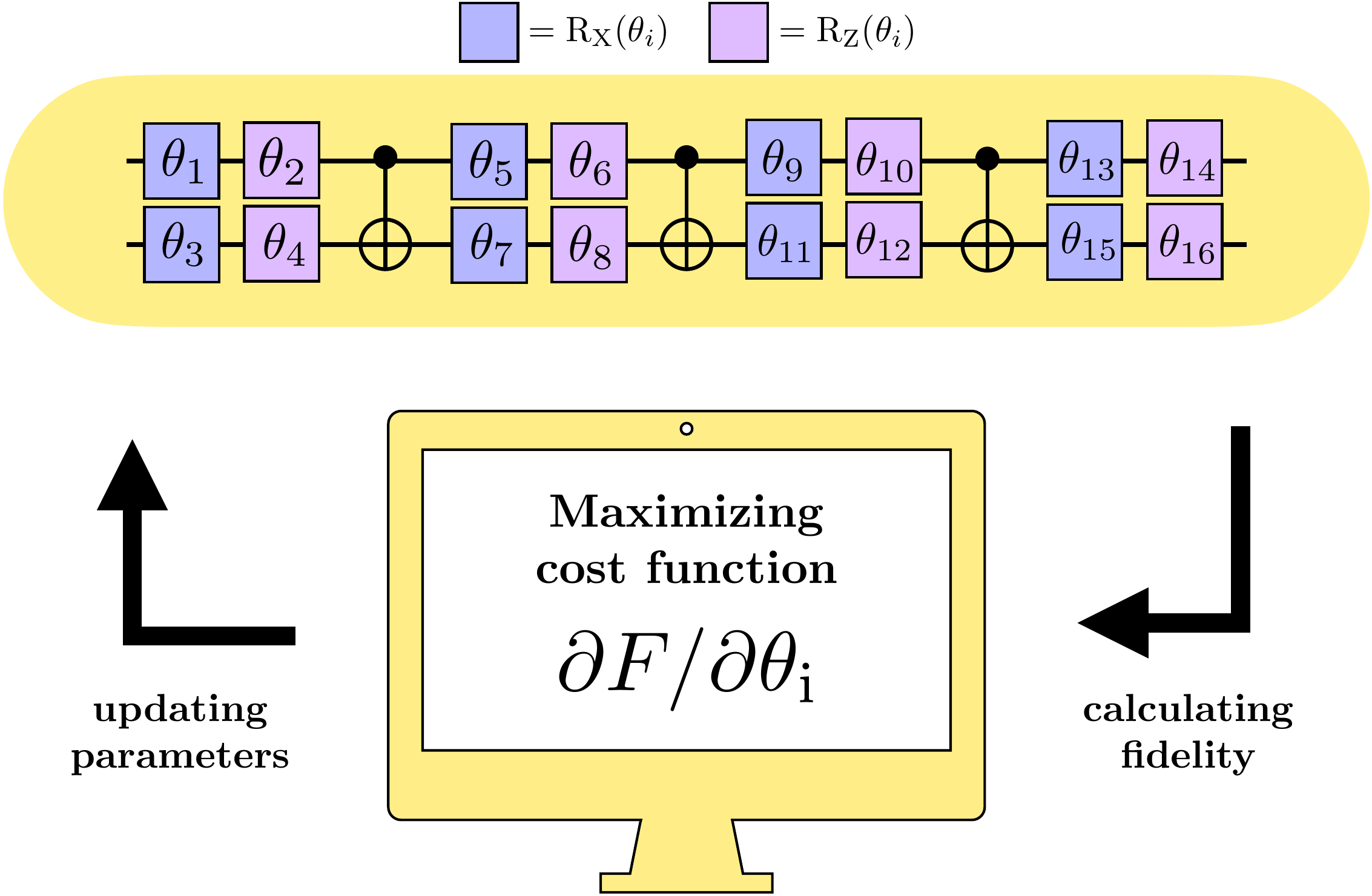}
\caption{\label{fig:hqc} 
A schematic view of the variational algorithm consisting of a quantum circuit constructed by tuneable parametric quantum gates of $R_X$ and $R_Y$ rotations, as well as CNOT gates. 
The circuit encircled in the yellow box (representing the unitary $U$) is employed by each party independently at their location.
Once each protocol step has been executed by the parties (including the measurements $M$), the cost function (i.e., the fidelity to the GHZ state) is computed. Subsequently, a classical gradient ascent algorithm is employed to calculate the derivatives of the fidelity with respect to the parameters.
The parameters are then updated according to the derivatives and the variational algorithm is run until the output fidelity converges to a final value.
}
\end{figure}

\section{Variational Algorithm}
An analytical approach to design a GHZ distillation protocol leads to heavy non-linear algebra. As an alternative, here
we utilise a VQA, a quantum-classical hybrid optimisation method, in order to efficiently determine the most appropriate $U$ matrix, and to manage the related computational cost.

In the VQA approach, a classical optimizer is used to train a parametric quantum circuit.
In our scheme, the parameterized quantum circuit is the circuit which realizes a single (or double) iteration of the protocol shown in Fig.~\ref{fig:setup}. The key element of the circuit is the ansatz for the unitary $U$, comprising of single-qubit parametric gates $R_X(\theta_{i})$ and $R_Z(\theta_{i})$, which rotate the quantum state around the $x$ and $z$ axes, respectively, and three two-qubit CNOT  gates in between blocks of the $R_{x}$ and $R_{z}$ gates, as encircled by the yellow box in Fig.~\ref{fig:hqc}. The ansatz includes measurements according to the required protocol configuration detailed in the previous section.
The structure of our ansatz is inspired by the circuit decomposition of a general two-qubit unitary operator in terms of single and two-qubit gates.
Three CNOT gates and sixteen single-qubit rotations are sufficient to decompose a general unitary operator for a two-qubit system \cite{Rakyta2022approaching}.

For the input of the parameterized circuit we choose a noisy GHZ state of the form 
\begin{equation}
    \rho_0=(1-\lambda)\rho_\textrm{GHZ} + \frac{\lambda}{8} I_8 \, ,
\end{equation}
where $\lambda$ denotes the noise strength and $I_8$ is the identity in the 3-qubit Hilbert space.

The optimization process (shown in Fig.~\ref{fig:hqc}) involves the following steps:
(1) Select random values for the gate parameters $\theta_i$ in the range $[0,2 \pi]$ and a random value for the noise strength $\lambda$ in the range $[0.05,0.3]$.
(2) Evaluate the circuit with the chosen parameters and determine the output state:
\begin{align}
    \rho_{ \{ \theta_i \}}=\mathcal{P}_{U\{ \theta_i \}}\left[\rho_0\right],
\end{align}
where $\mathcal{P}_U$ denotes the overall nonlinear transformation by the protocol.
(3) Evaluate the cost function which we define as the fidelity of the output state of the tailored circuit to the pure, noiseless GHZ state: 
\begin{align}
    \mathcal{F}=\langle \textrm{GHZ} |\rho_{ \{ \theta_i \}} | \textrm{GHZ} \rangle.
\end{align}
(4)  In order to find the new and improved set of parameters, we apply the Gradient Ascent Optimizer Method\cite{Goodfellow-et-al-2016}: The gradients of the fidelity 
are computed with respect to the parameters $\theta_i$ by 
the difference quotient formula
\begin{align}
    \frac{\partial\mathcal{F}}  {\partial\theta_i}=\frac{\mathcal{F}[\theta_i+\delta]-\mathcal{F}[\theta_i]}{\delta},
\end{align}
where the difference quotient $\delta$ was chosen to be equal to $0.01$.
(5) The parameters are updated according to the gradients,
\begin{align}
    \theta_{i+1}=\theta_{i}+\alpha \,\frac{\partial\mathcal{F}}{\partial\theta_i},
\end{align}
where the learning rate is $\alpha=0.1$.
(6) Randomly select a new value for the  noise strength $\lambda$ and reevaluate the circuit with the updated parameters. Stochastic selection of $\lambda$ values leads to a protocol that can successfully distill several input states with different noise strengths.
(7) Repeat the process iteratively until convergence is achieved, and the fidelity is effectively maximized.
Steps (1)-(6) are considered as a single optimization cycle.
We note that in our simulations $150$ optimization cycles were considered.

\section{Results}

In this section, we present two scenarios: (a) the cost function is computed after a single iteration, and (b) after two iterations of the protocol. 

\subsection{Single iteration protocol}

\begin{figure}[h]
\includegraphics[width=0.45\textwidth]{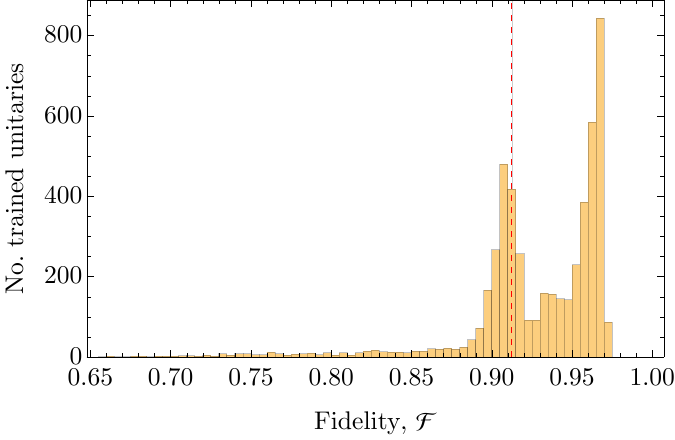}
\caption{\label{fig:his1} Histogram depicting the deviation of the output fidelities achieved by the unitary operators trained for a single iteration of the protocol by the VQA, which was executed 5000 times.
The unitaries were tested on the input state $\rho_\textrm{test}=(1-\lambda)\rho_\textrm{GHZ} + \lambda/8 I_8$, with $\lambda=0.1$, having an initial fidelity of $\mathcal{F}_0=0.9125$ to the GHZ state (denoted by the red dashed line).}
\end{figure}

In Fig. \ref{fig:his1} we show a histogram representing the output fidelities achieved by the unitaries optimized by our VQA algorithm. The VQA algorithm was executed $5000$ times for a single iteration of the protocol. Then, each unitary was applied on the input state $\rho_\textrm{test}=(1-\lambda)\rho_\textrm{GHZ} + \lambda/8 I_8$, with $\lambda=0.1$, having an initial fidelity of $\mathcal{F}_0=0.9125$ to the GHZ state. According to this test, the training was successful in $66\%$ of the events, resulting in a higher output fidelity than $\mathcal{F}_0$.
The top three operators $U^{s}_{j}$ leading to the highest output fidelities are shown in Table \ref{tab:pars1}, together with the corresponding $\theta_{i}$ parameters.

\begin{center}
\begin{table}[h]
\small
\begin{tabular}{ |c||c|c|c|c|c|c|c|c|  }
 \hline
  & $\theta_1$ & $\theta_2$ & $\theta_3$ & $\theta_4$ & $\theta_5$ & $\theta_6$ & $\theta_7$ & $\theta_8$ \\
 \hline
 $U_1^\textrm{s}$ & 1.59 & 0.29 & 2.69 & 2.77 & 0.05 & 2.63 & 2.72 & 0.11 \\
 \hline
 $U_2^\textrm{s}$ & 0.07 & 1.46 & 0.58 & 1.27 & 0.88 & 1.90 & 0.56 & 0.19 \\
 \hline
 $U_3^\textrm{s}$ & 2.97 & 1.92 & 3.03 & 1.50 & 3.13 & 1.28 & 1.60 & 2.26 \\
 \hline
 \hline
& $\theta_9$ & $\theta_{10}$ & $\theta_{11}$ & $\theta_{12}$ & $\theta_{13}$ & $\theta_{14}$ & $\theta_{15}$ & $\theta_{16}$ \\
 \hline
 $U_1^\textrm{s}$ & 1.65 & 0.54 & 1.25 & 1.24 & 3.14 & 3.03 & 1.87 & 3.14 \\
 \hline
 $U_2^\textrm{s}$ & 0.09 & 3.03 & 2.81 & 0.64 & 2.25 & 2.40 & 0.41 & 2.95 \\
 \hline
 $U_3^\textrm{s}$ & 0.16 & 1.04 & 0.11 & 0.02 & 0.00 & 2.95 & 1.61 & 3.02 \\
 \hline

\end{tabular}
\caption{\label{tab:pars1} The three best performing unitary operators and the corresponding optimized parameter values for the case when a single iteration of the protocol has been encoded into the cost function of the VQA, and the optimization procedure has been repeated a total of 5000 times.}
\end{table}
\end{center}

\begin{figure}[h]
\includegraphics[width=0.45\textwidth]{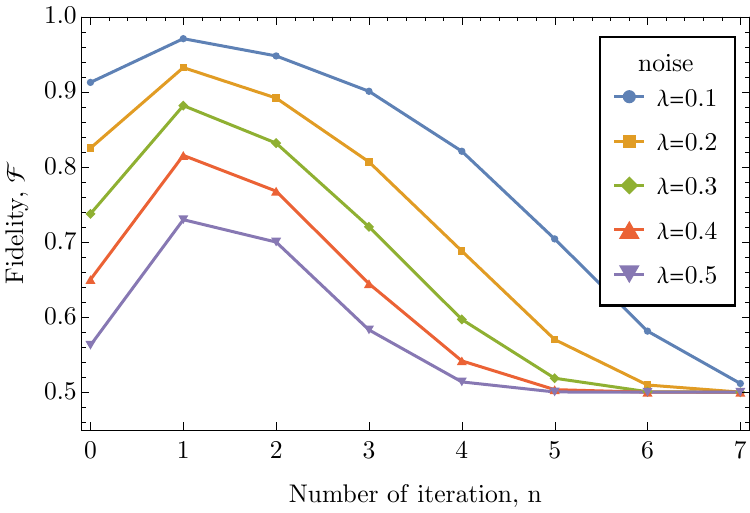}
\caption{\label{fig:fid-lam-1} 
The output fidelities obtained by the distillation protocol containing $U_{1}^{\textrm{s}}$ that gives the highest fidelity through optimization via the cost function for a single iteration of the protocol. 
The output fidelities are shown as function of the number of iterations for inputs of the form $(1-\lambda)\rho_\textrm{GHZ} + \lambda/8 I_8$, where $\lambda$ represents the noise strength. 
Lines connect the data points for ease of interpretation.
}
\end{figure}

In Fig.~\ref{fig:fid-lam-1} we analyze the performance of the operator $U_1^{\textrm{s}}$, corresponding to the highest fidelity ($\mathcal{F}_1=0.971$) achieved during the test, through various values of the noise strength $\lambda$. 
One can see that the protocol improves the fidelity by the first iteration, according to the way we constructed the cost function of the VQA, however, in the subsequent iterations, the fidelity begins to gradually decline. For example, for $\lambda=0.1$, after the seventh iteration the achieved state fidelity is only $\mathcal{F}_7=0.512$, meaning that we get significantly distant from a GHZ state.
Thus, in general, it is recommended to perform only one iteration with this unitary, as further iterations decrease the fidelity. 
We note that this is a generic behaviour of all other unitary operators obtained by the  same training that result in a high fidelity.

The unitary operators $U_i^{\textrm{s}}$ of Table~\ref{tab:pars1} have been trained to distill an initially mixed state. In what follows, we investigate 
whether these unitaries are capable of handling coherent errors of pure input states as well.
In order to do so, we consider two different pure-state models.

\begin{figure}[h]
\includegraphics[width=0.45\textwidth]{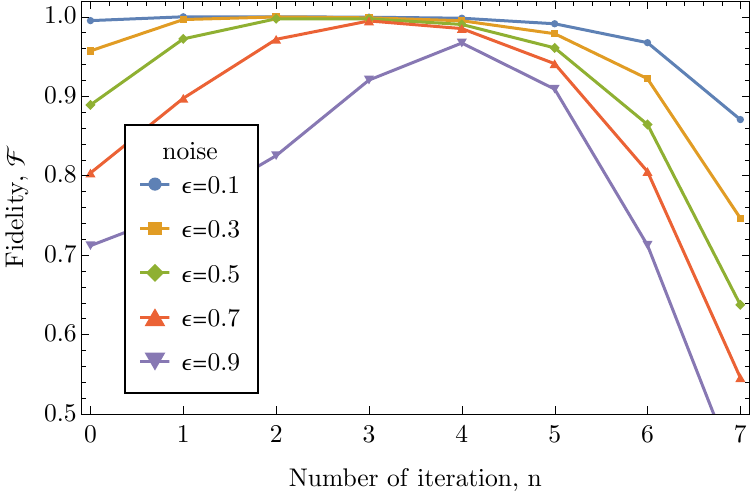}
\caption{\label{fig:fid-eps-1} 
The fidelity of the output state to the GHZ state as a function of the iteration number using the protocol containing $U_{1}^{\textrm{s}}$ for pure-state inputs of the form  $\mathcal{N}\left(|GHZ\rangle+\epsilon |011\rangle \right)$ for different values of $\epsilon$.
Points  are connected by lines for guiding the eye.
}
\end{figure}

In the first case, the efficiency of the protocol is investigated for inputs,  where the GHZ state is perturbed by an extra term of one of the computational basis states, e.g.,   
\begin{align}
    |\psi_0\rangle =\mathcal{N}\left(|GHZ\rangle+\epsilon |011\rangle \right),
    \label{eq:compst_err}
\end{align}
where we restrict ourselves to real values of the parameter $\epsilon$.
Figure~\ref{fig:fid-eps-1} shows the fidelities obtained by using $U_{1}^{\textrm{s}}$ as a function of the number of iterations for various values of  $\epsilon$. 
The protocol demonstrates its effectiveness, similarly to the case of mixed inputs, in enhancing the fidelity up to a certain number of iterations, beyond which the noise eventually becomes more pronounced again.
It can be seen that the protocol can deal with this type of error more efficiently than with white noise, as for $\epsilon<0.5$, after $2$-$3$ iterations, the state has a fidelity of $\mathcal{F}\sim0.999$ with the GHZ state. Below a threshold value of the noise parameter $\epsilon_\textrm{tresh}\approx1.01$ the protocol successfully distillates the GHZ state in a few iterations. 
We note that introducing any of the computational basis states ($|001\rangle,|010\rangle,|011\rangle,|100\rangle,|101\rangle,|110\rangle$) as the additional term in the input state of Eq.~(\ref{eq:compst_err}) would experience a similar effect.

\begin{figure}[]
\includegraphics[width=0.45\textwidth]{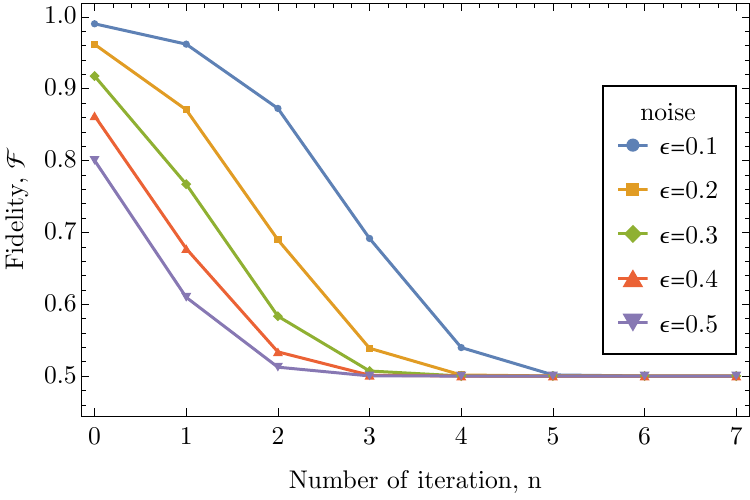}
\caption{\label{fig:fid-ghzl-1} 
The fidelity of the output state to the GHZ state as a function of the iteration number using the protocol containing $U_{1}^{\textrm{s}}$ for pure-state inputs of the form  $| \psi_0 \rangle=\mathcal{N}((1- \epsilon )|000\rangle + (1+\epsilon )|111\rangle)$ for different values of $\epsilon$.
Points  are connected by lines for guiding the eye.
}
\end{figure}

The second type of pure-state inputs we investigate are GHZ-like initial states of the form
\begin{align}
    | \psi_0 \rangle=\mathcal{N}((1- \epsilon )|000\rangle + (1+\epsilon )|111\rangle),
\end{align}
where we consider real values of $\epsilon$ in the range $[-1,1]$, representing the deviation of the state from the GHZ state. Figure~\ref{fig:fid-ghzl-1} shows the output fidelity as a function of the iteration number obtained by the protocol using $U_1^{\textrm{s}}$ for this type of inputs. We note that negative values of 
$\epsilon$ result in a similar behaviour. One can see that the protocol completely lacks the capability to correct this kind of error. Moreover, it can be shown that protocols, utilizing either of the $U_i^{\textrm{s}}$ operators would neither be able to correct this type of error. 

Since we have seen that the ability of these single-iteration protocols to correct incoherent as well as coherent errors is rather limited, in the next subsection we analyze whether building one more iteration into the cost function during training results in a better performance. 

\subsection{Double iteration protocol}

As demonstrated in the previous subsection, training the unitary operator by integrating a single iteration into the cost function during VQA optimization yielded an effective performance in the first iteration, but  experienced a decrease in the output fidelity in the subsequent iterations. In this subsection, we investigate whether training the unitary by including two iterations in the cost function can further enhance the preformance of the protocol.

\begin{figure}[]
\includegraphics[width=0.45\textwidth]{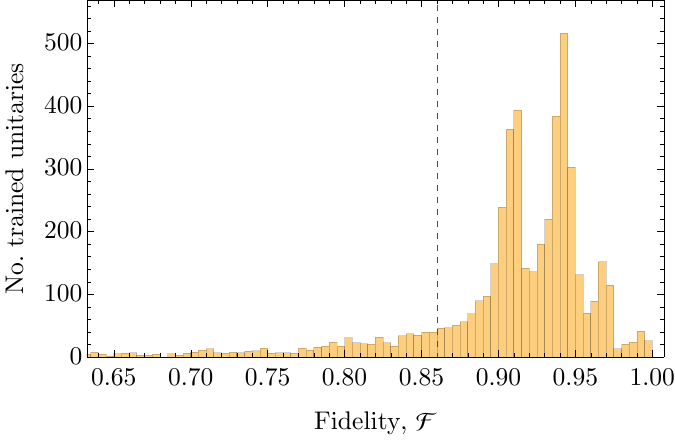}
\caption{\label{fig:his2} Histogram depicting the deviation of the output fidelities achieved by the unitary operators trained for a double iteration of the protocol by the VQA, which was executed 5000 times.
The unitaries were tested on the input state $\rho_\textrm{test}=(1-\lambda)\rho_\textrm{GHZ} + \lambda/8 I_8$, with $\lambda=0.1$, having an initial fidelity of $\mathcal{F}_0=0.9125$ to the GHZ state (denoted by the red dashed line).
}
\end{figure}

Figure~\ref{fig:his2} shows a histogram of the output fidelities achieved by the $5000$ unitaries obtained by the double-iteration VQA algorithm. The unitaries were tested on the same
$\rho_\textrm{test}$ input as in the single iteration case. 
According to the test, in $53\%$ of the cases the unitaries were successfully trained, resulting in output fidelities larger than the initial $\mathcal{F}_0=0.9125$.
The top three best performing operators, denoted as $U_j^\textrm{d}$ are listed in Table~\ref{tab:pars2} together with the respective $\theta_{i}$ parameters. By comparing Fig.~\ref{fig:his1} and Fig.~\ref{fig:his2}, one can see that the double-iteration case produces operators with significantly improved performance.
We note that the unitary denoted by $U_1^\textrm{d}$ in Table~\ref{tab:pars2} achieved the highest output fidelity of $\mathcal{F}_2=0.997$ out of all of the optimized unitaries on the $\rho_\textrm{test}$ data.
Thus, in what follows, we examine the performance of the protocol utilizing this operator.

\begin{figure}[]
\includegraphics[width=0.45\textwidth]{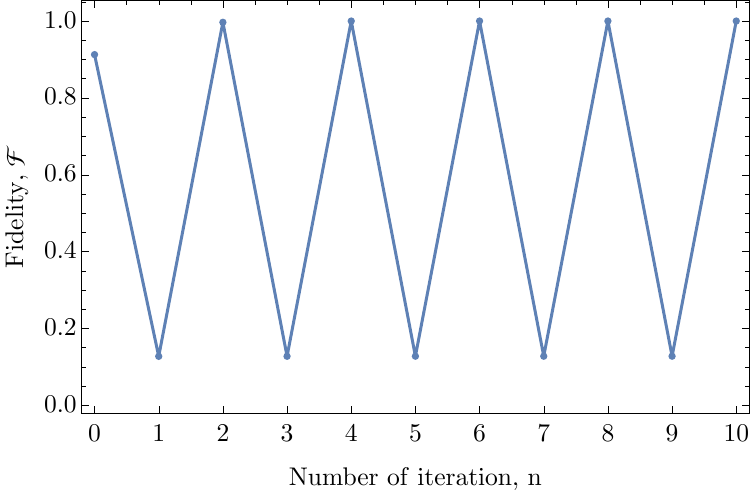}
\caption{\label{fig:fid2} 
The output fidelity as a function of the number of iterations for a protocol using $U_{1}^{\textrm{d}}$ that has been optimized by the VQA through the cost function corresponding to two iterations of the protocol. 
The input state is $(1-\lambda)\rho_\textrm{GHZ} + \lambda/8 I_8$, with $\lambda=0.1$. 
Data points are connected by lines to aid visualization.
}
\end{figure}

If we plot the output fidelity of the protocol using $U_{1}^{\textrm{d}}$ as a function of the number of iterations, as shown in Fig.~\ref{fig:fid2}, it can be seen that the fidelity oscillates between two values, indicating that there is a  convergence to a stable cycle of length two. For odd-numbered iterations the state is far from the desired GHZ state, however, for even-numbered iterations the GHZ state is distilled to a good approximation.
This behaviour is a consequence of the training criterion, as we imposed the cost function to increase the fidelity for a double iteration of the protocol.
We note that this is a generic feature of most of the operators trained this way.
Thus, if operators $U_i^\textrm{d}$ are used for distillation, the protocol should be iterated an even number of times.

\begin{center}
\begin{table}[]
\small
\begin{tabular}{ |c||c|c|c|c|c|c|c|c|  }
 \hline
  & $\theta_1$ & $\theta_2$ & $\theta_3$ & $\theta_4$ & $\theta_5$ & $\theta_6$ & $\theta_7$ & $\theta_8$ \\
 \hline
 $U_1^\textrm{d}$ & 3.10 & 1.83 & 0.00 & 1.13 & 0.56 & 1.67 & 3.05 & 1.56 \\
 \hline
 $U_2^\textrm{d}$ & 3.05 & 1.87 & 3.10 & 1.84 & 2.53 & 0.93 & 3.10 & 1.95 \\
 \hline
 $U_3^\textrm{d}$ & 1.49 & 0.87 & 3.07 & 1.21 & 1.56 & 1.56 & 2.28 & 1.50 \\
 \hline
 \hline
& $\theta_9$ & $\theta_{10}$ & $\theta_{11}$ & $\theta_{12}$ & $\theta_{13}$ & $\theta_{14}$ & $\theta_{15}$ & $\theta_{16}$ \\
 \hline
 $U_1^\textrm{d}$ & 3.14 & 2.37 & 3.04 & 2.49 & 1.81 & 2.90 & 0.20 & 3.11 \\
 \hline
 $U_2^\textrm{d}$ & 0.03 & 3.03 & 1.54 & 1.55 & 1.25 & 0.61 & 1.60 & 3.09 \\
 \hline
 $U_3^\textrm{d}$ & 3.14 & 0.77 & 2.37 & 1.63 & 2.08 & 1.51 & 2.46 & 2.97 \\
 \hline

\end{tabular}
\caption{\label{tab:pars2} The three best performing unitary operators and the corresponding optimized parameter values for the case when two iterations of the protocol were encoded into the cost function of the VQA. The optimization procedure has been repeated a total of 5000 times.
}
\end{table}
\end{center}

\begin{figure}[]
\includegraphics[width=0.45\textwidth]{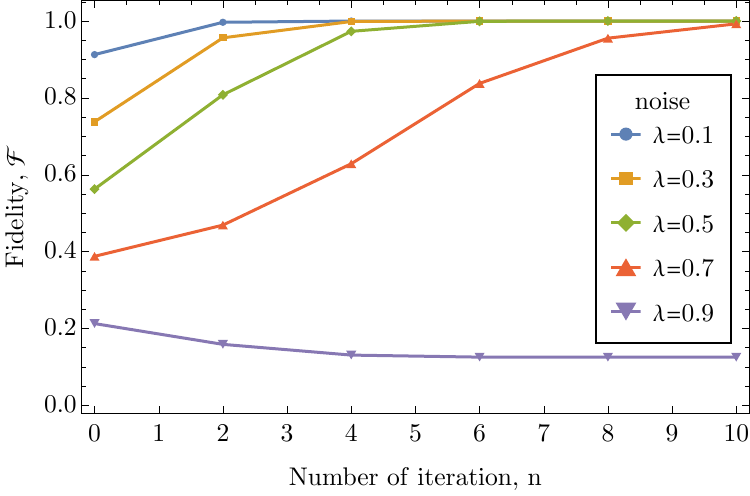}
\caption{\label{fig:fid-lam-2}
The output fidelity as a function of the number of iterations for the protocol using $U_{1}^{\textrm{d}}$, for inputs of the form $(1-\lambda)\rho_\textrm{GHZ} + \lambda/8 I_8$, and for different values of $\lambda$. Note that only even-numbered iterations are shown and data points are connected by lines.
}
\end{figure}

In order to analyze the tolerance of the distillation protocol utilizing $U_1^\textrm{d}$, we employed different levels of white noise strength $\lambda$ and calculated the output fidelities for even-numbered iterations, as shown in  Fig.~\ref{fig:fid-lam-2}. 
It can be seen that above a critical value of the parameter $\lambda_\textrm{crit}\approx0.79$, the protocol effectively improves the output fidelity. For $\lambda\leq 0.3$, the protocol achieves a close-to-perfect fidelity of $\mathcal{F}\approx1$ by the fourth iteration.
It is worth noting that with the use of $U_1^\textrm{d}$, a continuous fidelity improvement can be achieved (provided the noise is less than the critical value $\lambda_{c}$), whereas with $U_1^\textrm{s}$ the fidelity only improved in the first iteration, after which it gradually decreased. Thus, utilizing $U_1^\textrm{d}$ in the protocol allows ones to perform GHZ distillation within a few iterations with high fidelity.

In what follows, we investigate whether the protocol containing $U_{1}^{\textrm{d}}$ can be used to transform distorted pure states to the GHZ state.
As a reminder, we note that $U_i^\textrm{d}$ has been trained to distill mixed states, therefore it is not trivial whether it can be used to correct coherent errors as well. We apply the same models of errors as in the previous section. 

\begin{figure}[]
\includegraphics[width=0.45\textwidth]{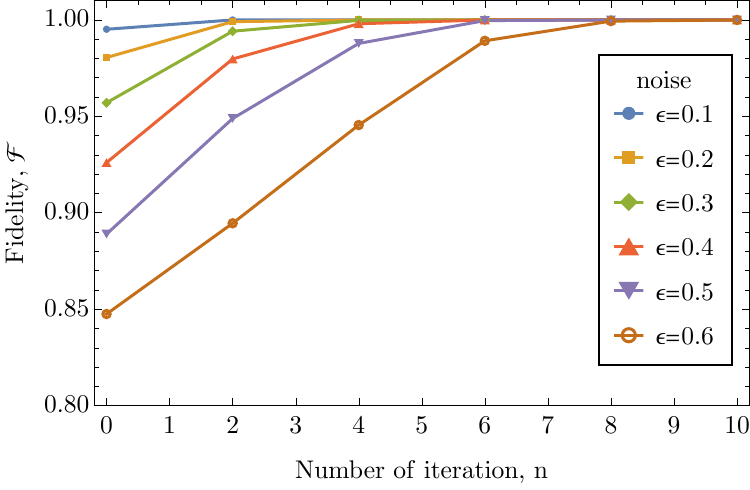}
\caption{\label{fig:fid-eps-2}
The output fidelity achieved by the protocol utilizing $U_1^\textrm{d}$ as a function of even-numbered iterations, for pure-state inputs with coherent error represented as $\mathcal{N}\left(|GHZ\rangle+\epsilon |011\rangle \right)$.
Points  are connected by lines for guiding the eye.}
\end{figure}

Let us first consider initial pure states of the form $|\psi_0\rangle=\mathcal{N}\left(|GHZ\rangle+\epsilon |011\rangle \right)$, which  deviate from the GHZ state by an additional term (one of the computational basis states), the deviation being characterized by the real parameter $\epsilon$. The achieved output fidelities are plotted in Fig.~\ref{fig:fid-eps-2}, as a function of the iteration number for different values of $\epsilon$.
One can see that the protocol demonstrates an effective improvement in the fidelity by a few, even-numbered iterations. 
We note that including any of the computational basis states ($|001\rangle,|010\rangle,|011\rangle,|100\rangle,|101\rangle,|110\rangle$) as a coherent error in the input, would lead to a similar result. Thus, this type of error can be corrected by the protocol.

\begin{figure}[]
\includegraphics[width=0.45\textwidth]{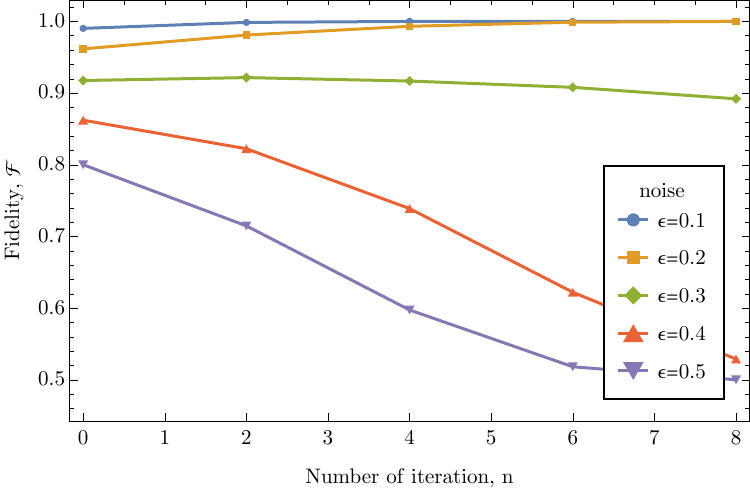}
\caption{\label{fig:fid-ghzl-2} 
The output fidelity achieved by the protocol utilizing $U_1^\textrm{d}$ as a function of even-numbered iterations, for pure-state inputs with coherent error represented as $| \psi_0 \rangle=\mathcal{N}((1- \epsilon )|000\rangle + (1+\epsilon )|111\rangle)$. Points  are connected by lines for guiding the eye.
}
\end{figure}

Let us now consider GHZ-like pure input states 
$| \psi_0 \rangle=\mathcal{N}((1- \epsilon )|000\rangle + (1+\epsilon )|111\rangle)$, where $\epsilon$ denotes the coherent deviation from the GHZ state. As shown in Fig.~\ref{fig:fid-ghzl-2}, $U_1^\textrm{d}$ is effective to remove this type of error for smaller values of $\epsilon$ (namely, $\epsilon < 0.3$), in contrast to $U_1^\textrm{s}$, which is never successful.

We can conclude that the protocol using $U_1^\textrm{d}$ is more effective than the one using $U_1^\textrm{s}$, 
as $U_1^\textrm{d}$ consistently increases the fidelity (and eventually reaches significantly higher) by every even-numbered iteration in the case of mixed initial states, unlike $U_1^\textrm{s}$ which only improves the fidelity in the first iteration. Similarly, for pure, but coherently distorted initial states, $U_1^\textrm{d}$ successfully removes the error for small distortions in contrast to $U_1^\textrm{s}$ which can deal with this error only in a few specific cases.

The fact that the double iteration protocol provides a better distillation scheme might suggest that longer iteration protocols could outperform the ones presented in this work, however, our analysis shows that increasing the number of iterations does not result in a faster convergence to the GHZ state.

\section{Conclusion}

We have presented quantum protocols involving unitary transformations acting on two qubits and subsequent measurements and post-selection, optimized for the purpose of distilling GHZ states in an LOCC scheme. We employed variational optimization methods to train the unitaries, in order to prepare outputs with high fidelity to the GHZ state after a few iterations. Especially, the case when the target of optimization was the fidelity to the GHZ state after two iterations of the same protocol led to surprisingly effective schemes. In this case, we found an even-odd oscillation of the output fidelity. We found that considering only an even number of iterations, the fidelity increases rapidly and approaches unity after a few iterations for a rather large class of input states, containing either coherent deviations from the GHZ state or white noise. Our results indicate that optimizing a fixed number of iterations of a circuit with less inputs can provide a way to find effective distillation protocols. We note that in the optimization procedure we used a general ansatz containing three CNOT gates. One might also look for less general circuits containing a smaller number of CNOT gates, which are expected to lead to a slower convergence.

\section*{Acknowledgements}

This research was supported
by the Ministry of Culture and Innovation, and  the National Research, Development and Innovation Office within the Quantum Information National Laboratory of Hungary (Grant No. 2022-2.1.1-NL-2022-00004), the KDP-2021 scheme (Grant No. C1790232), the ÚNKP-23-5 New National Excellence Program, by the J\'{a}nos Bolyai Research Scholarship of the Hungarian Academy of Science and by
the NKFIH through the OTKA Grants FK 132146 and
FK 134437. TK and OK were supported by the NKFIH (TKP-2021-NVA-04).

\bibliographystyle{elsarticle-num}
\bibliography{paper}

\end{sloppypar}
\end{document}